# Ultra-fast magnetization manipulation using single femtosecond light and hot-electrons pulse

Y. Xu[1,2], M. Deb*[1], G. Malinowski[1], M. Hehn,[1] W. Zhao[2], S. Mangin[1]

1) Institut Jean Lamour, CNRS UMR 7198, Université de Lorraine, 54506 Vandœuvre-lès-Nancy, France
2) Fert Beijing Research Institute, BDBC Beihang University, Beijing 100191, China

**Abstract:**

Current induced magnetization manipulation is a key issue for spintronic application. Therefore, deterministic switching of the magnetization at the picoseconds timescale with a single electronic pulse represents a major step towards the future developments of ultrafast spintronic. Here, we have studied the ultrafast magnetization dynamics in engineered $Gd_x[FeCo]_{1-x}$ based structure to compare the effect of femtosecond laser and hot-electrons pulses. We demonstrate that a single femtosecond hot-electrons pulse allows a deterministic magnetization reversal in either Gd-rich and FeCo-rich alloys similarly to a femtosecond laser pulse. In addition, we show that the limiting factor of such manipulation for perpendicular magnetized films arises from the multi-domain formation due to dipolar interaction. By performing time resolved measurements under various field, we demonstrate that the same magnetization dynamics is observed for both light and hot-electrons excitation and that the full magnetization reversal take place within 5 ps. The energy efficiency of the ultra-fast current induced magnetization manipulation is optimized thanks to the ballistic transport of hot-electrons before reaching the GdFeCo magnetic layer.

*Correspondence to: M. Deb ([Marwan.deb@univ-lorraine.fr](Marwan.deb@univ-lorraine.fr))

.

Energy efficient and ultrafast magnetization manipulation without any applied magnetic field is of utmost importance for both future spintronic application and fundamental understanding of ultrafast magnetization dynamics. Indeed the dynamic response of magnetic order to ultrafast external excitation is a fascinating issue of modern magnetism[1]. Moreover an efficient mean of magnetization manipulation is of prime importance for information and memory storage devices. During the last ten years the effect of spin-transfer torque (SST)[2–5], spin-orbit torque[6,7], electric field[8–10] and femto- or picosecond lasers pulses[11–13] have motivated numerous theoretical and experimental investigations. The recent discoveries in the field of all optical switching have attracted a lot of attention especially the All Optical Helicity Independent Switching (AO-HIS) of GdFeCo using a single femtosecond laser pulse. The exact microscopic mechanisms still need to be understood, but both experimental and theoretical results indicate the pure thermal origin of AO-HIS in GdFeCo[11,14,15]. Concerning applications, the magnetization switches within 30 ps which allows pushing the speed of the recording technology to high frequencies above 10 GHz[16]. However in the context of spintronic devices, the magnetization switching induced by electronic currents and electric field are particularly interesting as they can be implemented in nanoscale devices[17,18].

The technological importance of the ultrafast AO-HIS in GdFeCo immediately raised the question of the possibility to obtain the same reversal speed with a single electron pulse, which can be adapted for nanoscale technologies and open the field of ultrafast-spintronic. Hot-electrons transport has been recently presented as an alternative way to manipulate spins and has been demonstrated to offer an extensive control of the ultrafast demagnetization in metallic multilayers[19], as initially demonstrated in Ni film by femtosecond laser pulses[20]. Furthermore, spin polarized hot-electrons allows controlling the speed of demagnetization[21] and even initiate coherent magnetization precession by STT[22]. More recently, Williams *et al.* have shown the possibility to reverse the magnetization by hot-electrons in GdFeCo[23]. However, the timescale of such a magnetization reversal remains unknown. In addition, this magnetization reversal is only reported for one composition of GdFeCo. These results triggered different fundamental questions: What exactly is the timescale of this switching process? Is it similar to the one reported for the light induced AOS? Since the demagnetization timescale of rare-earth and transition metal magnetic atoms is different, can hot-electrons reverse the magnetization in both Gd-rich and FeCo-rich GdFeCo alloys? If yes, does the two reversal timescales are the same? What is the hot-electrons transport mechanism?

In order to answer these questions, we have carefully studied the magnetization dynamics induced in $Gd_x[FeCo]_{1-x}$ by hot-electrons and laser pulses as a function of the film composition, laser fluence, and external magnetic field (see suplementary material). In addition to switching induced by direct excitation of the laser pulse, we evidence that a single hot-electrons pulse reverses the magnetization in both Gd-rich and FeCo-rich samples without requiring any additional magnetic field. Furthermore, we demonstrate that the magnetization reversal process takes in a few picoseconds and that thanks to a ballistic hot-electrons transport ultra-short electrons pulse allows energy efficient magnetization manipulation.

**I. EXPERIMENTAL METHOD AND STATIC SAMPLE PROPERTIES:**

Our study is based on a series of 5 nm thick $Gd_x[FeCo]_{1-x}$ layers made of an amorphous alloy with different concentrations of Gd ranging from 20 to 30 %. The alloys are ferrimagnetic metallic materials, having two antiferromagnetically exchange coupled sublattices, formed by the magnetic moments of the transition metals $Fe_{86}Co_{14}$ (FeCo in the following) and the rare earth Gd. Therefore, by tuning the composition of the $Gd_x[FeCo]_{1-x}$ it is possible to tune the magnetic properties, which paves the way for a general understanding of the physical phenomena. The investigated samples were deposited by dc magnetron sputtering onto a glass substrate according to the following artificial multilayer structure: Glass/Ta (3 nm)/Pt (5 nm)/ Cu (80 nm)/$Gd_x[FeCo]_{1-x}$ (5 nm)/Ta (5 nm). The thin Ta capping layer prevents the oxidation of the magnetic film, while allowing probing the magnetic properties via the magneto-optical Kerr effect (MOKE).

Figure 1(a) shows the composition dependence of the coercive field ($H_c$) and the saturation magnetization ($M_s$) in the 5 nm $Gd_x[FeCo]_{1-x}$ films, which are respectively measured at room temperature (RT) via MOKE in polar configuration and superconducting quantum interference device-vibrating sample magnetometer (SQUID-VSM). For all compositions and with the field applied perpendicular to film plane, the films have a square hysteresis loop revealing a perpendicular magnetic anisotropy and so a magnetization oriented perpendicular to the film plane. At room temperature, the net magnetization ($M_s$) reaches zero at x=26%, which corresponds to the compensation concentration ($x_{comp}$) of the system, where the magnetization of both Gd ($M_{Gd}$) and transition metal sublattices ($M_{FeCo}$) are equal. Therefore, we show as expected at $x_{comp}$ a clear divergence of $H_C$. $M_s$ is thus aligned in the direction of the CoFe moments below $x_{comp}$ while it changes its sign and becomes aligned with the Gd moments above $x_{comp}$. This is in agreement with the sign inversion observed between the hysteresis loops for Gd-rich and CoFe-rich [inset of Fig.1(a)] measured via the Kerr rotation ($\Theta_K$), since $\Theta_K$ is mainly sensitive to the FeCo moments in the visible light ($\lambda$= 628 nm)[24,25]. Let us mention that with increasing the temperature of Gd-rich sample, $M_s$ vanishes at the so-called compensation temperature ($T_M$), while in CoFe-rich sample $T_M$ is below the RT [24].

Using our engineered multilayer structures combined with femtosecond laser pulses, we can investigate the ultrafast magnetization dynamics induced by two types of ultrashort excitations: (i) An optical excitation as the laser pulse interacts directly with the magnetic film when the sample is irradiated from the front side [Fig. 1(b)] (ii) An indirect excitation induced only by hot-electrons, which can be performed by an optical pumping of the sample from the back side [Fig. 1(c)]. In this configuration, the magnetic film is buried under an 80 nm thick Cu layer, leading to a very low transmission at 800 nm (less than 0.1%)[19]. The ultrafast magnetization dynamics induced separately by femtosecond laser and hot-electrons pulses were investigated by the pump-probe configurations sketched in the Fig. 1(b) and Fig. 1(c), respectively. The pump and probe beams are generated from a 35 fs laser pulses delivered by a Ti:sapphire regenerative amplifier operating at 800 nm with a repletion rate of 5 KHz. The pump beam (800 nm) excites the sample at normal incidence, while the probe beam (400 nm) has a small angle of incidence of about 5°. They are both linearly polarized and focused onto the sample. The spot diameters of the pump are ~260 μm and ~330 μm for the back and the front sides of the sample, respectively, while the spot diameter of the probe is 60 μm. During the pump probe measurements, a static external magnetic field $H_{ext}$ of 1T is applied perpendicularly to the film plane, which allows saturating the samples and therefore reset the magnetic state before every laser pulse. The differential Kerr rotation $\Theta_K$ signals ($\Delta\Theta_K$) are detected with a polarization bridge and synchronous detection scheme as a function of the delay between the pump and the probe. A static Kerr microscope is also used to observe the magnetic domains induced after the excitation of the sample by 800 nm femtosecond laser pulses with a spot size of ~50 μm.

II. RESULTS AND DISCUSSION:
II.1 ULTRA-FAST MAGNETIZATION DYNAMICS FOR A BROAD RANGE OF CONCENTRATIONS:

Figure 2(a) displays the MOKE images obtained after one, then two, three and four direct light excitations of $Gd_x[FeCo]_{1-x}$ films by femtosecond laser pulses. A single shot helicity independent all optical switching (AO-HIS) is observed in samples with an amount of Gd (23<x<28) corresponding to concentration above and below the compensation concentration ($x_{comp}$ = 26). However, for the concentration far from $x_{comp}$, a multi-domain state is obtained (x = 20.5). This result tend to show that the single pulse (AO-HIS) can be observed only if the magnetic domain size is larger than the light spot, as it has been shown for multiple pulse All Optical helicity dependent switching (AO-HDS)[26]. This criteria is fulfilled for very thin films and/or low saturation magnetization ($M_s$) which can be obtained near $x_{comp}$[26,27]. This observation suggests that AO-HIS could be observed for a larger range of composition in the case of a smaller laser spot or a thinner film. A more detailed analysis is proposed in the supplementary material ( see Figure S1). The images obtained for various laser fluences (fig. 2a) prove that samples having a composition closer to compensation require less power to switch. Moreover, the influence of the laser fluence on the switching confirms the existence of a power threshold which depends on the concentration.

Figure 2b displays the MOKE images obtained after one, then two, three and four electron pulses excitation generated by a laser pulse of difference fluences. Again Both Gd-rich samples and FeCo-rich samples show clear single pulse switching. From the comparison between Fig. 2a and 2b we can conclude that a similar behavior is observed when the sample is excited with a hot-electrons pulse or with a direct laser pulse. Even the evolution as a function of the laser fluence is qualitatively similar for both excitations. Compared to the direct excitation, the laser fluence needed to create the hot electron pulse has to be 4 times higher than for the direct light pulse to observe similar switching. In the supplementary material Figure S1 present the comparison of the MOKE images obtained for light and electron pulses for six different concentrations.

For a better understanding of the magnetization reversal mechanism, time-resolved pump-probe technique was employed to monitor the magnetization dynamics with a temporal resolution lower than 50 fs. The time-resolved MOKE (TR-MOKE) measurements of the magnetization dynamics induced by laser and hot-electrons pulses in $Gd_xFeCo_{1-x}$ films are displayed in Fig. 3(a) and Fig. 3(b), respectively. The amplitude of $H_{ext}$ is adjusted to be slightly larger than $H_C$ for each sample in order to ensure the same initial magnetic state. Let us first focus on the magnetization dynamics induced by the laser pulses. An ultrafast demagnetization occurring within 1 ps is observed for all samples. After this ultrafast demagnetization, we distinguish three different behaviors depending on the composition of $Gd_x[FeCo]_{1-x}$. The first is characterized by a rapid relaxation of the magnetization back to the initial state as can been seen for the sample with $x_{Gd}$=20.5%, which has a high $M_s$ and didn't show AO-HIS using MOKE (Fig. 2a). The second occurs for samples with moderate $M_s$ ($x_{Gd}$=22.5% and 28.6%) which have shown multi-domain state (Fig. 2a). It is characterized by a long lived fully demagnetized state. The third behavior is the switching toward negative values which is observed for the compositions near $x_{comp}$, where the samples are characterized by low $M_s$ and consequently large magnetic domains. As it will be demonstrated the first behavior is observed when the Zeeman energy is large and that consequently magnetization tend to follow the applied field and relax back to its initial state. The second is observed when the magnetic layer tends to brake into small domains due to the dipolar field leading to a zero net magnetization. The third behavior is a clear demonstration of single pulse AO-HIS which is obtained for both CoFe-rich and Gd-rich samples and confirm the previous MOKE images. For both type of concentrations it occurs qualitatively on the same time scale. Indeed, after a full demagnetization within ~1 ps, the magnetic state shows a slow variation until ~3.2 ps, then a high speed reversal process is initiated until reaching almost 100% reversal within ~40 ps. All the previous steps are confirmed and presented in the supplementary material figure S4 by measuring the hysteresis loops at different delay times . A clear sign inversion of $\Theta_K$ is observed after a time delay of 5 ps. The TR-MOKE induced by hot-electrons is displayed in Fig. 3b. Like for the optical excitation, a sub 5ps magnetization reversal is induced by hot-electrons for samples with a composition near $x_{comp}$. In addition, for the concentration far from $x_{comp}$, a long lived full demagnetization state or a rapid relaxation toward the initial state are observed depending on the sample. Let us mention that for the sample with $x_{Gd}$ of 22.2%, showing single pulse AO-HIS in the static Kerr microscopy images, only a long lived full demagnetized state is obtained in TR-MOKE measurements. The main difference between the two measurements is the pump spot size. The large spot size in the TR-MOKE setup makes more likely the formation of magnetic domains and could explain the origin of the difference[26]. Figure 3c and 3d show the power dependence of TR-MOKE signal measured 100 ps after excitation of the sample by laser and hot-electrons pulse. For both excitations, three different behaviors can clearly be distinguished as a function of the laser fluence. First, a recovery of the magnetization toward its initial state is shown at low fluence for all samples ($\Delta\Theta_K/\Theta_K$ at 100 ps ≈1). Second, a long lived partial demagnetization state is also observed for all samples (1>$\Delta\Theta_K/\Theta_K$ at 100 ps >0). Finally, an AO-HIS is obtained above a threshold fluence within a certain composition range around $x_{comp}$ (22<x<28). At a very high fluence, the signal measured for samples showing AO-HIS decreases (Fig. 3d, $x_{Gd}$= 25.6%). This is due to the creation of a multidomain state, which is located in the center of the pump beam as directly observed by Kerr microscopy. Those conclusions are in accord with the magnetization dynamic as a function of the magnetic field amplitude presented in the supplementary materials. Finally, note that the value of laser fluence cannot be directly related to the energy absorbed by the GdFeCo films. Indeed, for example in the hot-electrons configuration, the Cu (80nm) layer reflects more than 85% of the incident light.

To further evaluate the importance of the magnetization switching by hot-electrons for nanoscale spintronic purposes, it is interesting to compare their characteristics with those obtained in the STT-MRAM technology. To determine the intrinsic hot-electrons switching energy, $E_S$, we have used the transfer matrix calculation to estimate the absorption profile of the sample[1]. We find that only ~12% of the light is absorbed by the Pt/Cu layer. To determine the switching energy in different structure size, one can use the following formula which is adapted from ref[28]:

$$E_s = FA\alpha$$

Where F is the switching threshold fluence, A is the switched area, and α is the absorption coefficient at the pump wavelength. For a switching threshold fluence of ~10 mJ.cm$^{-2}$, we estimate a hot-electrons switching energy $E_S$ of ~4 fJ for an area of $20 \times 20$ nm$^2$. This energy value is one order of magnitude lower than the one required in STT-MRAM memories[29,30]. In addition, the switching speed obtained by hot-electrons is in the picosecond timescale, which is about thousand times faster than in the case of STT-MRAM[29,30].

**II.2 Cu THIKNESS DEPENDENCE OF THE ULTRAFST MAGNETIZATION DYNAMICS:**

To reveal the mechanisms behind magnetization switching using hot-electrons, the ultrafast magnetization dynamics induced by hot electrons in Glass/Ta (3 nm)/Pt (5 nm)/ Cu ($t_{Cu}$)/Gd$_{24.5}$[FeCo]$_{75.5}$ (5 nm)/Ta (5 nm) films are studied as a function of the Cu thickness ($t_{Cu}$) which is varied from 5 to 200 nm. In order to determine the zero time delay with high accuracy, each sample is patterned in alternating areas with and without Cu, which gives the time for which direct excitation occurs. As $t_{Cu}$ increases, the laser fluence needs to be increased to generate hot electrons and reverse the magnetization due to an increase in the sample reflection. To compare the magnetization dynamics for all the samples, we adjusted the laser fluence to obtain the same demagnetization amplitude as in the case of direct laser excitation for a fluence of 3.2mJ.cm$^{-2}$(Fig. 3a). TR-MOKE measurements as a function of $t_{Cu}$ are displayed in Fig. 4a. Two important features can be identified. First, a full demagnetization is reached within 2 ps for $t_{Cu}$ =200 nm, which is totally opaque to the light.

This clearly establishes that hot-electrons transport is at the origin of the observed magnetization dynamics. We also emphasize that, for all Cu thickness, hot-electrons can reverse the magnetization of GdFeCo films. This result clearly confirms that the hot-electrons only are reversing the magnetization. Second, the onset time of the magnetization dynamics continuously increases as $t_{Cu}$ increases. To determine the characteristic properties of the demagnetization process as a function of Cu thickness, we adjusted the TR-MOKE signals with the following empirical model:

$$\Delta\Theta_K(t)/\Theta_K = \{A_1 - A_2/[1 + e^{[(t-(t_0+2.2\tau))/\tau]}]\} + A_2$$

Where $A_1$ is the amplitude of the signal at negative time delay, $A_2$ is the value reached at maximum demagnetization, $t_0$ is the time corresponding to 10 % of the maximum demagnetization, and $\tau$ is the demagnetization characteristic time. The results of this fits allow obtaining the relative time delay $\Delta t_0=[t_0(t_{Cu})-t_0(d=5nm)]$ and the modification of the characteristic demagnetization time $\Delta\tau=[\tau(d)-\tau(d=5nm)]$, which are displayed respectively in Fig. 4.b and Fig. 4.c. To compare with literature, we plot also in these figures the Cu thickness dependence of the induced $\Delta t_0$ and $\Delta\tau$ in the ultrafast demagnetization of Cu($t_{Cu}$)/[Co/Pt](4nm) multilayers[19]. One observes that $\Delta t_0$ increases linearly with the Cu thickness. The slope of the linear fit yield a velocity v= 0.53 $10^6$ m/s, which is of the same order of magnitude as found in Cu ($t_{Cu}$)/[Co/Pt] (~0.68 $10^6$ m/s). The high velocity value is a direct imprint of hot-electrons ballistic transport. Furthermore, the ultrafast demagnetization in Cu($t_{Cu}$)/[ Co/Pt] can be reproduce only by theoretical simulation based on hot-electrons ballistic transport[19]. These two properties rule out the contribution of pure thermal transport in the ultrafast process induced by hot-electrons. Furthermore, we show that $\Delta\tau$ increases linearly with Cu thickness. This can principally be related to the broadening of the hot-electrons pulse. On the other hand, the corresponding increases of $\Delta\tau$ is found more pronounced in GdFeCo than Co/Pt multilayers. One possible reason for this difference is that the demagnetization process induced by hot-electrons in the localized 4f moment in Gd is slower than in 3d magnetic moment in Co. Another reason could be that the hot-electrons attenuation within the ferromagnetic layer, which can be slightly higher in GdFeCo with thickness of 5nm than in Co/Pt multilayers with a total thickness of 4 nm.

### III. CONCLUSION:

In conclusion, our results clearly demonstrate that deterministic magnetization manipulation in thick GdFeCo films can be observed using either a single femto-second laser pulse or a single femto-second hot-electrons pulse without requiring any magnetic field. In both cases the effect is shown for a broad range of GdFeCo compositions above and below the compensation. One of the limiting factors for the observation of the magnetization switching is the presence of the stray field that generates multi-domain state in the case of large magnetization. Importantly we prove by studying the time-resolved magnetization dynamics that the reversal process induced by hot-electrons and light are similar and take place within 5 ps for both Gd-rich and FeCo-rich films. We suggest that the similar magnetization dynamics obtained by both type of excitations is related to the ballistic nature of hot-electrons transport which we clearly demonstrated in our especially engineer structure. Finally, we are convinced that such a fast magnetization switching using electron pulse opens the door to ultra-fast spintronic.

## METHODS

The AO-HDS measurements were performed using optical pulses, having a central wavelength of 800 nm (1.55 eV), a pulse duration of about 35 fs at the sample position and a 5 kHz repetition rate. The response of the magnetic film was studied using a static Kerr microscope to image the magnetic domains after the laser excitation. The present measurements are performed at room temperature (RT).

**ACKNOWLEDGMENTS:** The authors thanks C. Chang, M.S. El Hadri, T. Fache, E.E. Fullerton, T. Hauet, G. Kichin, G. Lengaigne, F. Montaigne, Y. Quessab, P. Scheid, P. Vallobra, and B. Zhang for fruit-full discussion and technical help. The authors gratefully acknowledge the financial support of the Agence National de la recherche in France via the projects COMAG: # ANR-13-IS04-0008-01 and UMANI: # ANR-15-CE24-0009 and the International Collaboration Project B16001 and National Natural Science Foundation of China (Grant No. 61627813). Experiments were carried out on IJL Project TUBE-Davms equipments funded by FEDER (EU), PIA (Programme Investissemnet d'Avenir), Region Grand Est, Metropole Grand Nancy and ICEEL.

**AUTHOR CONTRIBUTIONS:** M. D, M.H, G.M and S.M., designed and coordinated the project; Y.X grew, characterized and optimized the samples. Y.X. M. D and G.M designed and operated the Faraday microscope and the pump laser set-up. M.D and S.M coordinated work on the paper with contributions from Y.X, M.H, G.M and W.Z and regular discussions with all authors

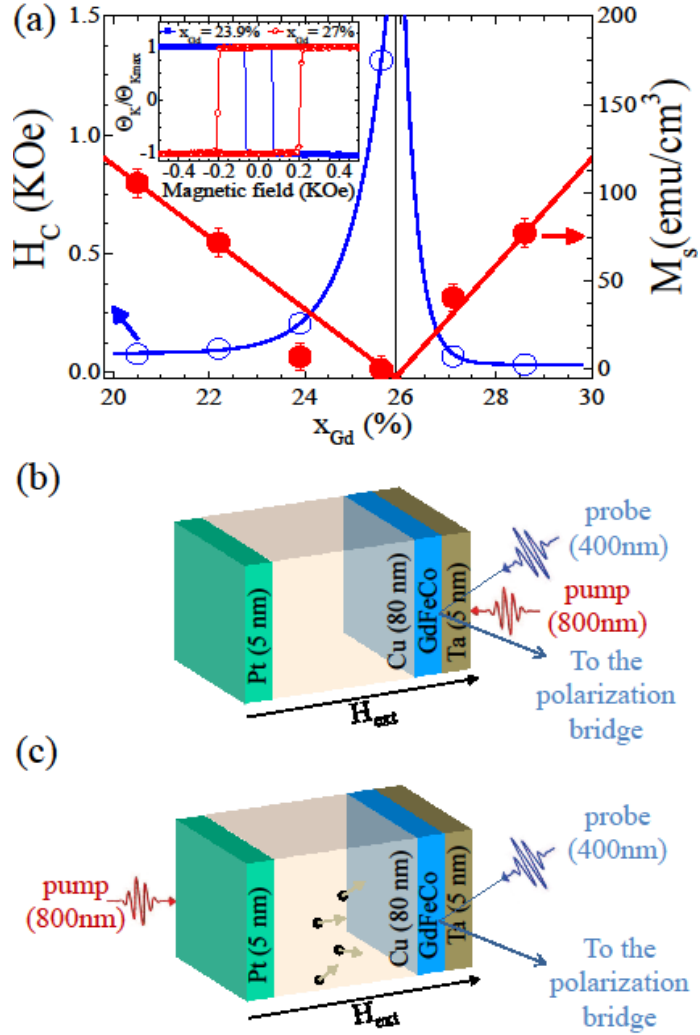

Figure 1: Quasi-static room-temperature magnetic properties of Glass/Ta (3 nm)/Pt (5 nm)/ Cu (80 nm)/$Gd_x[FeCo]_{1-x}$ (5 nm)/Ta (5 nm) films for (20 %< $x_{Gd}$ <30%) and the pump-probe experimental configurations. (a) Variation of the coercive field $H_C$ and the saturation magnetization $M_s$ as a function of the Gd content ($x_{Gd}$). The inset shows a normalized polar hysteresis loops for the films with $x_{Gd}$ of 23.9% and 27%. (b,c) Sketch of the time resolved experimental configurations allowing to separately study the ultrafast magnetization dynamics induced by single light pulse and hot-electron pulse.

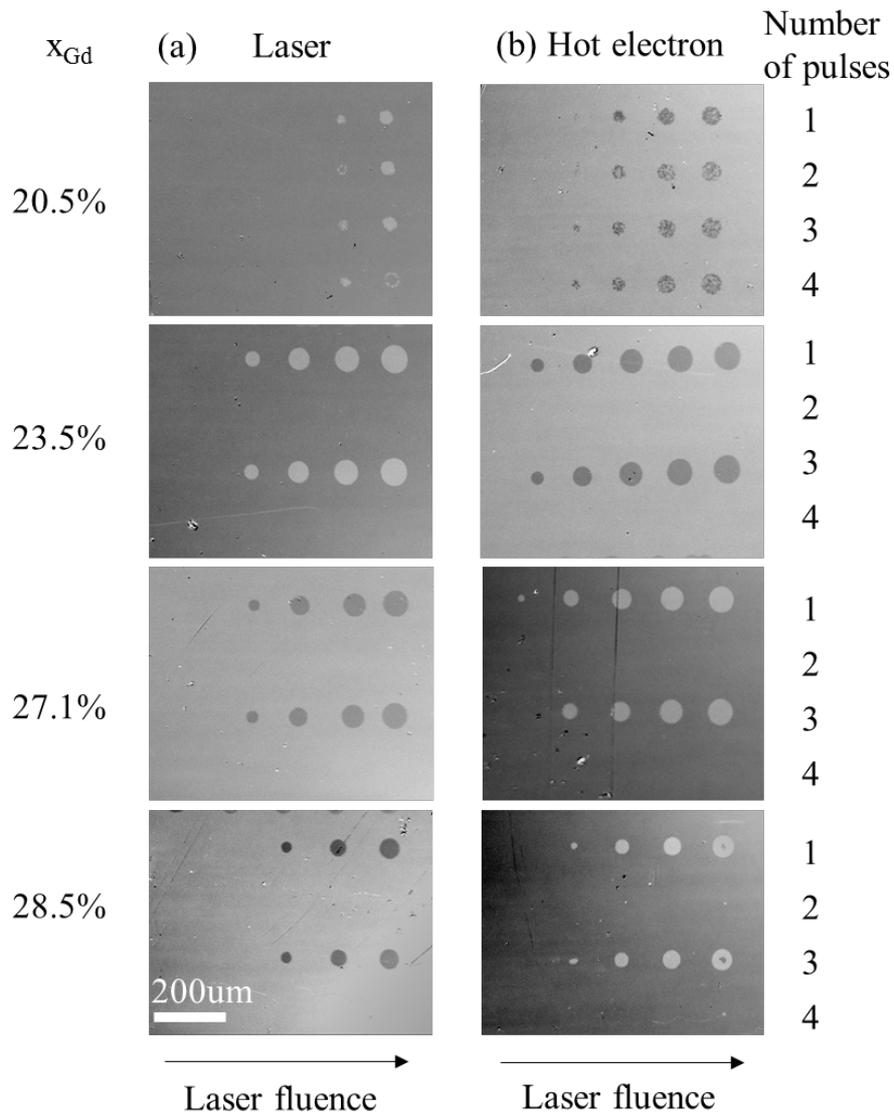

Figure 2: Magneto-optical Kerr images obtained after the excitation of $Gd_x(FeCo)_{1-x}$ films by (a) one then two, three and four direct light pulses for five different values of laser fluences ranging from 0.75 to 3.5 mJ.cm$^{-2}$ and (b) one then two, three and four hot-electron pulses for five different values of laser fluences ranging from 4 to 12.25 mJ.cm$^{-2}$.

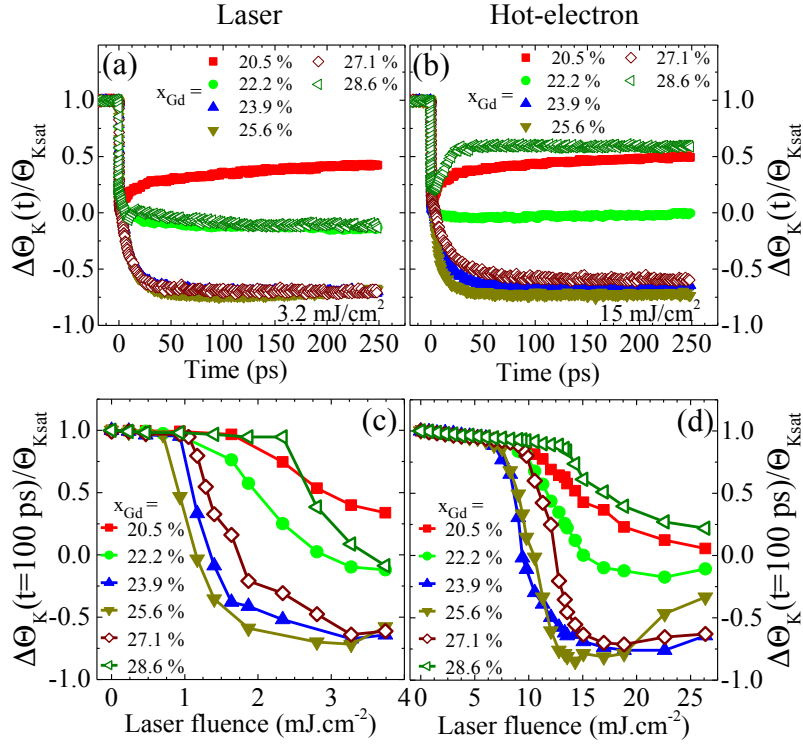

Figure 3: Comparison between the magnetization dynamics induced in 5 nm thick $Gd_x(FeCo)_{1-x}$ films by femtosecond laser and hot-electron pulses for six different concentration above and below the compensation concentration. (a,b) $\Delta\Theta_K(t)/\Theta_{Ksat}$ induced by laser (a) and hot-electrons (b) pulses after excitation of the samples with fixed laser fluencies of 3.2 mJ/cm$^2$ and 15 mJ/cm$^2$, respectively. (c, d) Laser fluence dependence of the $\Delta\Theta_K/\Theta_{Ksat}$ signal measured at 100 ps after excitation of samples by laser (c) and hot-electron (d) pulses.

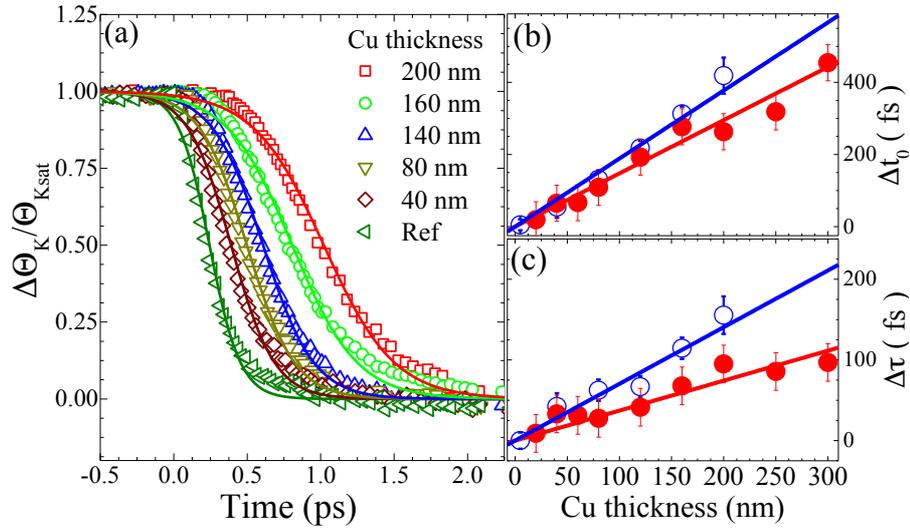

Figure 4: Magnetization dynamics in Glass/Ta (3 nm)/Pt (5 nm)/ Cu ($t_{Cu}$)/Gd$_{24.5}$[FeCo]$_{75.5}$ (5 nm)/Ta (5 nm) films measured as a function of Cu thickness (Glass/Ta (3 nm)/Pt (5 nm)/ Cu ($t_{Cu}$)/Gd$_{24.5}$[FeCo]$_{75.5}$ (5 nm)/Ta (5 nm)). (a) $\Delta\Theta_K(t)/\Theta_{Ksat}$ measured as a function of Cu thickness. The solid line is a fit using Eq. (1). (b,c) Comparison of the Cu thickness dependence of $\Delta t_0$ (b) and $\Delta\tau$ (c) obtained in this work (open circles) and those reported for the ultrafast demagnetization in Cu($t_{Cu}$)/[ Co/Pt] multilayers (adapted from ref. [19]). The line is a linear fit to the data.